\documentclass{jpsj3}
\usepackage{txfonts}

\title{Interlayer Conductance in the Armchair Nanotube - Zigzag
Graphene Ribbon Parallel Contact: Theoretical Proposal of Detection of Wavefunction Growing from the Edge to the Center in the Graphene Ribbon} 
\author{Ryo Tamura}
\inst{Faculty of Engineering, Shizuoka University, 3-5-1 Johoku, Hamamatsu 432-8561, Japan}

\abst{
Sublattices A and B are opposite in the decay direction of the edge state of the zigzag graphene ribbon (ZGR). Detecting exponential {\it growth} from the zigzag edges to the ZGR center remains challenging. 
The tight-binding model calculations in this letter reveal that interlayer conductance manifests this growth in parallel contact with the armchair nanotube. The transfer integrals of oblique interlayer bonds are comparable to those of vertical interlayer bonds. However, the phase 
of the ZGR wave function strongly suppresses the contribution of oblique bonds, allowing the selective detection of the growing component. }

\begin{document}
\maketitle

Since their discovery \cite{Iijima}, carbon nanotubes (NT) have attracted significant attention owing to their electronic and mechanical 
properties, accompanied by high aspect ratios. \cite{NT-22-review,2010-tamura-paper-NT-general}
These characteristics
are suitable as tips of atomic force microscopy, \cite{NT-5-SWNT-AFMtip,NT-6-SWNT-AFMtip,NT-tuika-SWNT-AFMtip} Kelvin force
microscopy \cite{NT-7-MWNT-Kelvein-tip,NT-tuika-MWNT-Kelvein-tip}, electrical probes
\cite{NT-8-electrical-probes, NT-9-electrical-probes}, and scanning tunneling microscopy (STM)
\cite{NT-10-SWNT-STM-tip,NT-11-metal-coated-tip,NT-12-metal-coated-tip}; the nanometer-sized radius guarantees a high spatial resolution.
Using the chiral index $(n_1,n_2)$, the NTs were metallic when
$n_1-n_2$ was a multiple of three (mod($n_1-n_2,3)$=0) and semiconducting otherwise (mod($n_1-n_2,3)$=1,2).\cite{NT-22-review,2010-tamura-paper-NT-general}. 
Progress continues in single-chirality separation.\cite{NT-1-separation,NT-2-separation,NT-3-separation}

An important target of the NT tip measurement is the edge states of the zigzag graphene ribbon (ZGR) \cite{zig-9-10-discovery}, which has been evaluated from various aspects: spin Seebeck effects \cite{zig-9-spin-Seebeck}, magnetism
\cite{zig-9-12-discovery-magnetism,zig-9-9-magnetism,zig-9-13-magnetism,zig-9-14-magnetism}, spin transport \cite{zig-10-spin-transport,zig-11-spin-transport}, valleytronics \cite{a4-3.,40.,41.,a4.}, and zero-conductance dips \cite{zig-12-dip,zig-13-dip}.
STM signals of the edge states appear at the zigzag edge \cite{zig-1-only-exp,zig-2-only-exp,zig-3-exp-theory,zig-4-exp-theory,zig-7-defect} but are absent at the armchair edge \cite{zig-5-armchair,zig-6-armchair}.
In a standard STM setup, the NT tip is
perpendicular to the ZGR surface, whereas the NT $\pi$ orbitals are orthogonal to those of the ZGR.
Contrarily, theoretical calculations show that the $\pi$-orbital mixing governs the $I$-$V$ characteristics \cite{NT-18-NTGr-cond-theory,NT-19-NTGr-cond-theory} and charge distribution 
\cite{NT-17-charge-transfer-D-theory} in the parallel contacted graphene-NT junctions.
Experimental studies on this parallel setup have been reported regarding the conductance \cite{NT-13-NTGr-cond-exp, NT-14-NTGr-cond-exp, NT-15-NTGr-impedence-exp} and charge transfer. \cite{NT-16-charge-transfer-exp}
We can slide an NT with weak interlayer cohesion when the axial contact length is short.
The interlayer distance remains almost constant, similar to that in the constant-height mode of the STM.
However, the contact area is larger than the standard
STM tip, and the current variation with the slide motion
reflects the atomistic information.
Armchair NTs (ANTs) suit this measurement, allowing
a simple interlayer registration.
The molecular dynamic simulations confirmed the following:
The ANT axis tends to be parallel to the zigzag edge because AB stacking was the most stable. \cite{2019-NT-MD-1,2019-NT-MD-2-interlayer-distance,2019-NT-MD-3,2019-NT-MD-4}
In addition to the theoretical studies mentioned above, 
interlayer vibrations \cite{NT-4-MD-NT-Gr-resonator}
and spectral functions \cite{NT-21-spectral-function} have been studied theoretically. 
However, surveys of the edge state with this parallel ANT tip have yet to be reported.

\begin{figure}
\begin{center}
\includegraphics[width=0.9\linewidth]{17985Fig1xyz.eps}
\caption{(color online) (a)Geometric structure of the $(n,n)$ ANT and $(n',n')$ ZGR in the calculation. 
The atomic $y$ coordinates are $y=al/2$ with integers $l$
and the lattice constant $a=0.246$ nm.
The ZGR is on the $xy$ plane with the AB sublattice structure
and the outermost sites
are the B sites on the left $(x=0)$ and A sites on the right ($x=(3n'-1)a_{\rm c}$), 
where $a_{\rm c}=a/\sqrt{3}$.
The ANT atomic $x$ and $z$ coordinates are 
$x=R\sin\theta_{l,j}+(M-0.5)a_{\rm c}$ and $z=R(\cos\theta_{l,j}-1)-D$,
where $\theta_{l,j}=\frac{\pi}{n} (j-\frac{(-1)^j}{6}-\frac{(-1)^l}{2})$, $R=\frac{n\sqrt{3}a}{2\pi}$,
$D=$ 0.31 nm with integers $j$ and $M$.
(b)Atomic $(x,y)$ coordinates
in case $M=5$ and $M=13$.
The dotted lines are the ANT covalent bonds that face
the ZGR $(\cos\theta_{l,j} > 0)$, and the ovals indicate those nearest
to the ZGR. We only consider
cases $M=3m+2$ (configuration A) and $M=3m+1$ 
(configuration B) with $m$ integers. 
In case $M=3m$, the interlayer configuration becomes unstable AA stacking. }
\end{center}
\end{figure}

In this study, we discuss the $(n,n)$ armchair nanotube (ANT) that
partially overlaps $(n',n')$ ZGR as shown in Fig. 1 (a).
In ZGR and ANT, the atomic $y$ coordinates are $y=al/2$ with an integer $l$ and a lattice constant $a=0.246$ nm.
ZGR is on the $xy$ plane $(z=0)$ with an AB sublattice structure and the outermost sites are the B sites on the left $(x=0)$ and A sites on the right ($x=(3n'-1)a_{\rm c}$), 
where $a_{\rm c}=a/\sqrt{3}$ is the covalent bond length.
Although the ANT has an AB sublattice structure, 
we use sublattice symbols $A$ and $B$ only for the ZGR.
The ANT atomic $x$ and $z$ coordinates are 
$x=R\sin\theta_{l,j}+(M-0.5)a_{\rm c}$ and $z=R(\cos\theta_{l,j}-1)-D$, where $\theta_{l,j}$ denotes the angle $\frac{\pi}{n} (j-\frac{(-1)^j}{6}-\frac{(-1)^l}{2})$ with an integer $j$. 
 $R=\frac{n\sqrt{3}a}{2\pi}$ is the tube radius.
$(M-0.5)a_{\rm c}$ denotes the distance between the ZGR left edge and the ANT axis.  $D $ denotes the interlayer distance. 
In this definition of $\theta_{l,j}$, the ANT covalent bond is parallel to ZGR when nearest to it.
According to Refs. \cite{NT-17-charge-transfer-D-theory,2019-NT-MD-2-interlayer-distance,2021-sc-NT-0.31nm} , we choose $D=$ 0.31 nm. 
The ANT and ZGR are semi-infinite in the $y$-direction with armchair edge terminations at $y=\frac{a}{2}(N-2)$ and $y=0$, followed by the overlapped length $\frac{a}{2}(N-2)$ with an integer $N$.
According to the abovementioned atomic positions,
we define the tight-binding Hamiltonian as in Refs. \cite{Lambin, Lambin-2} .
The intralayer elements equal $-t$ ($=-2.75$ eV) for the nearest neighbors, and zero otherwise.
The interlayer elements become nonzero only when the atomic distance $\overline{r}$ is shorter than the cutoff distance $r_c=$ 0.39 nm.
The nonzero element is defined as 
$ t_1\exp[(r_1-\overline{r})/r_2]\cos(\theta_{l,j})$ 
with parameters $t_1=$ 0.36 eV, $r_1= $ 0.334 nm, and $r_2=$ 0.045 nm.
The exact numerical calculations were performed in a manner similar to that described in Ref. \cite{2019-Tamura} .

Figure 1 (b) illustrates the atomic $(x,y)$ coordinates
in cases $M=5$ and $M=13$.
Hereafter, $n=5$ and $n'=40$.
The dotted lines represent the ANT covalent bonds that face
ZGR $(\cos\theta_{l,j} > 0)$, whereas the ovals indicate the nearest
to the ZGR.
Because the stable interlayer configuration is
AB stacking, we considered only the
cases $M=3m+2$ and $M=3m+1$ with
an integer $m$. 
Interlayer bonds normal to the ZGR
correspond to the maximum interlayer transfer integral and are referred to as 'vertical' bonds here.
In this case, the vertical bonds are limited to sites A and B,
$M=3m+2$ and $M=3m+1$, respectively.
Hereafter, these cases are referred to as configurations A and B.
Without loss of generality,
we chose the range $m \leq 20$, where $m=20$
corresponds to the center of ZGR (vertical dashed line in Fig. 1 (a)).

\begin{figure}
\begin{center}
\includegraphics[width=\linewidth]{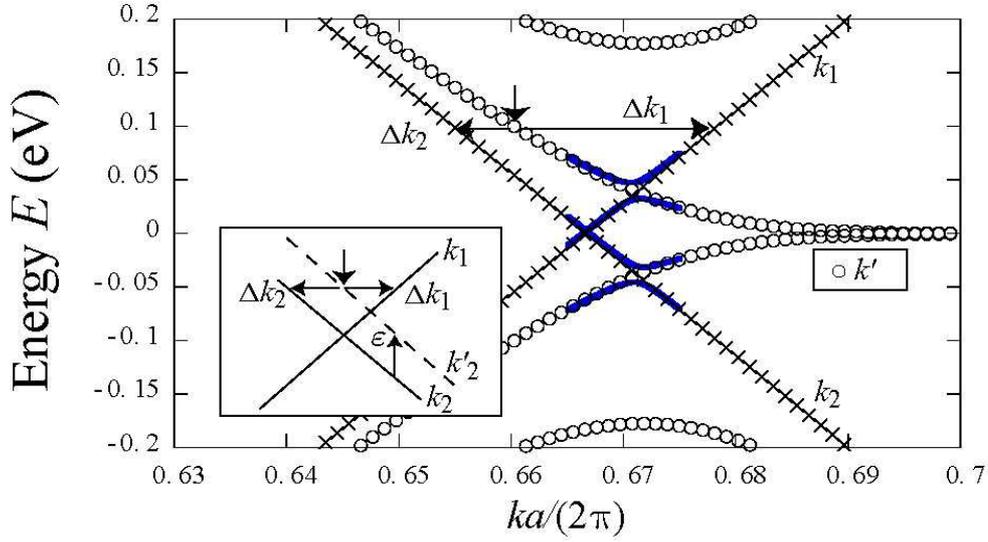}
\caption{(color online) The main panel is the dispersion relations
of the present system.
The x marks (circles) show the wave number $k_1,k_2$ $(k')$ 
of the isolate (5,5) ANT (the isolate (40,40) ZGR).
The solid lines represent the wave number $k_{\rm int}$ 
with the interlayer Hamiltonian in case $M=16$ (configuration B).
The horizontal arrows represent $\Delta k_\tau=k_\tau-k'$ in Eq. (\ref{PF})
when $E=$ 0.1 eV.
Inset is a schematic view of the linear 
$k$ and $k'$ dispersion lines
of {\it isolate} $(n,n)$ and $(n',n')$ ANTs
with the inter-tube site energy
difference $\varepsilon$.
For succinctness, $k'_1$ is omitted in the inset. 
}
\end{center}
\end{figure}

Reference \cite{2021-Tamura} shows the perturbation formula (PF) of the interlayer transmission rate $T_{\tau,\tau'} $ from channel $\tau'$ of the $(n',n')$-ANT to channel $\tau$ of the $(n,n)$-ANT for side-contacting ANTs (sc-ANTs).
The first-order PF is determined by the perturbation Hamiltonian and zeroth order wave function i.e., interlayer Hamiltonian elements 
and electronic states of the {\it isolated } ANTs.
In our notation, $ \exp(ik_\tau a l/2)c_{[l],j}^{\;(\tau)}$ denotes the wave function amplitude at site ($l,j$)
of isolate $(n,n)$ ANTs, where $c_{[l],j}^{\;(\tau)}$ is real and $k_\tau$
is the wave number with the channel index $\tau$. 
When $l$ is odd (even), $[l] =1$ ($[l] =2$).
The inset in Fig. 2 shows a schematic of the linear 
$k$ and $k'$ dispersion lines of the $(n,n)$ and $(n',n')$ ANTs where $k'_1$ is omitted for simplicity.
The encapsulated dopants induce the intertube-site energy
difference, $\varepsilon$.

In the case of Ref. \cite{2021-Tamura} , the relations
\begin{equation}
\frac{dk_\tau}{dE} \simeq \frac{2}{\sqrt{3}ta},
\label{dk/dE}
\end{equation} 
\begin{equation}
\frac{dk' }{dE} \simeq \frac{2}{\sqrt{3}ta},
\label{dk'/dE}
\end{equation} 
\begin{equation}
\Delta k_1 \simeq\frac{4E-2\varepsilon}{\sqrt{3}ta}
\label{dk1}
\end{equation} 
\begin{equation}
\Delta k_2 \simeq \frac{2\varepsilon}{\sqrt{3}ta}
\label{dk2}
\end{equation} 
hold and enables us to rewrite the PF as
\begin{equation}
T_{\tau} =\frac{Y_\tau}{\Delta k_\tau^2}
\sin^2\left(\frac{\Delta k_\tau}{4}aN\right)
\cos^2\left(\frac{\Delta \widetilde{k}_\tau}{4}aN+\sigma\frac{\pi}{3}\right)
\label{PF}
\end{equation}
where $\Delta k_\tau=k_\tau-k'$, $\Delta \widetilde{k}_1=\Delta k_2, \Delta \widetilde{k}_2=\Delta k_1,
\sigma={\rm mod}(N,3)$, 
\begin{equation}
Y_{\tau} =\left|\frac{dk_\tau}{dE}\right|\left|\frac{dk'}{dE}\right|
\frac{Z_\tau^2}{X_\tau},
\label{Y}
\end{equation}
\begin{eqnarray}
Z_{\tau} 
&=&\sum_{l=1}^2\sum_{\Delta l =-1}^1
\exp\left(i\frac{2}{3}\pi\Delta l\right)
\langle H \rangle_\tau^{l,l+\Delta l} 
\nonumber \\
&=&\sum_{l=1}^2\sum_{l'=1}^2
(-1)^{l-l'}
\langle H \rangle_\tau^{l,l'},
\label{Z}
\end{eqnarray}
\begin{equation}
\langle H \rangle_\tau^{l,l'} = 
\sum_{j'=1}^{2n'}\sum_{j=1}^{2n}
c_{[l],j}^{(\tau)}H^{(l,l')}_{j,j'}
c'_{[l'],j'},
\end{equation}
and 
\begin{equation}
X_\tau
= 
\sum_{l'=1}^2\sum_{j'=1}^{2n'}|c'_{l',j'}|^2
\sum_{l=1}^2\sum_{j=1}^{2n}|c_{l,j}^{(\tau)}|^2.
\label{X}
\end{equation}
Here, we fixed and suppressed the channel index $\tau'$ of the $(n',n')$ ANT
and $H^{(l,l')}_{j,j'}$ indicates the interlayer Hamiltonian elements between sites $(l,j)$ and $(l',j')$.
The factor $(-1)^{l-l'}$
originates from the approximation $ka \simeq k'a \simeq 4\pi/3$
and strongly suppresses the effect of the oblique interlayer bonds.

In the PF of the proposed system, $k'$
and $c'$ are the wave number and wave function
of the isolate ZGR, respectively.
The ZGR sites are labeled using integer indices $(l',j')$ 
in the same manner as in the ANT.
The PF is calculated using the correct $k$ and $k'$
except for the $(-1)^{l-l'}$ factor mentioned above.
In contrast to the ANT, the ZGR $k'$ dispersion line 
is highly nonlinear and $\frac{dk'}{dE}$, $\Delta k_1$ and
$\Delta k_2$ differ from those in Eq. (\ref{dk'/dE}), (\ref{dk1}), and (\ref{dk2}).
\begin{figure}
\begin{center}
\includegraphics[width=\linewidth]{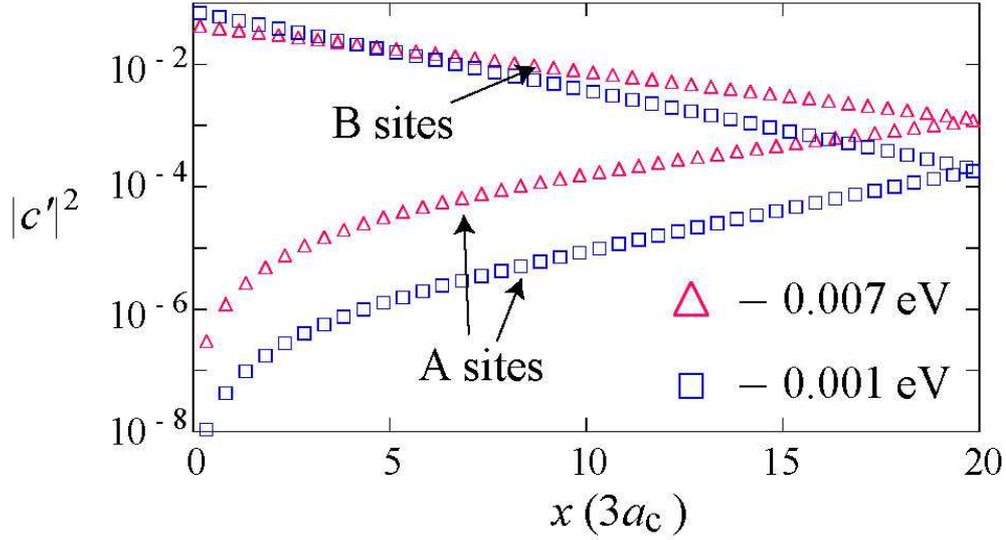}
\caption{(color online) Squared wave function amplitude $|c'|^2$ 
of the isolate (40,40) ZGR as a function of the $x$ position in the $3a_{\rm c}$ unit.
The NT axis comes near $x=3a_{\rm c} m$ with the ANT $x$ position index $m$
. The ZGR center corresponds to $m=20$.
}
\end{center}
\end{figure}
The main panel of Fig. 2 illustrates the dispersion relations of the present system.
The x marks indicate the wave numbers $k_1$ and $k_2$ of the isolate (5,5) ANT and are the same as those in the inset.
In contrast, the circles correspond to the isolates (40,40) ZGR
and represent a single dispersion line for the edge state near zero $E$. 
In Fig. 2, the horizontal arrows represent $\Delta k_\tau$ 
measured using $k'$.
Although there was no interlayer site energy difference,
in the proposed system, the difference between the linear and flat bands work effectively as $E$-dependent $\varepsilon$.
Because our focus is the zigzag edge state, we focused on the energy region $|E| < 0.177$ eV, in which only a single edge channel exists in the ZGR asymptotic region $ y > (N-2)a/2$.
This energy range is inversely proportional to $n'$.

The zeroth order wave function of ANT is quite simple because $c_{l,j}^{(2)}=(-1)^j$ and $c_{l,j}^{(1)}=1$. 
That of ZGR ($c'_{l,j}$) is relatively complicated, and is documented in Ref. \cite{zig-8-zeroth-edge} .
Figure 3 shows the squared amplitude $|c'_{l,j}|^2$ as a function of the $x$ position when $E=-0.007$ eV and $E=-0.001$ eV.
The B-site amplitude was localized at the left edge and decayed exponentially with $x$.
Conversely, the A-site amplitude increased exponentially and coincided with the amplitude of the B-site at the center, $x=60a_{\rm c}$.
As $|E|$ increases, the edge state becomes delocalized.

The solid lines in the main panel of Fig. 2 represent the wave number $k_{\rm int}$ where the interlayer Hamiltonian corresponds to the scattering region $0 \leq y \leq (N-2)a/2$
in case $M=16$ (configuration B).
Owing to the ANT curvature, the interlayer bonds only appear when $\cos\theta_{l,j} \simeq 1$. 
This curvature effect causes the $k_{\rm int}$ dispersion lines
extremely close to $k$ and $k'$
except for the narrow energy gaps at the $k$ - $k'$ cross (the solid lines are displayed only near the crossing).
Because of the B-site localization, the gap is wider in configuration B than in configuration A when the ANT is near the left edge.
For example, the gap regions are
$-46 $ meV $ <E < -31$ meV and 33 meV $<E<$ 47 meV in case
$M=16$ (Fig. 2),
and $-38 $ meV $ <E < -37$ meV and 38 meV $<E<$ 40 meV
for $M=17$ (not shown in the figures). 
Equations (\ref{dk'/dE}), (\ref{dk1}), and (\ref{dk2})
does not hold for the present system, as previously mentioned.
The PF is calculated using $k'$
and $c'$ of the isolate ZGR.
In contrast to sc-ANT, whether we can prove the PF of the present system is yet to be determined.
However, it should be noted that Eq.  (\ref{Y}) is a factor of Fermi's 
golden rule.\cite{2010-Tamura}
By comparing Eq. (\ref{PF}) with the exact transmission rates, we evaluate the effectiveness of Eq. (\ref{PF}).
The same tight-binding Hamiltonian was applied to both the
exact calculation and perturbation formula.

\begin{figure}
\begin{center}
\includegraphics[width=\linewidth]{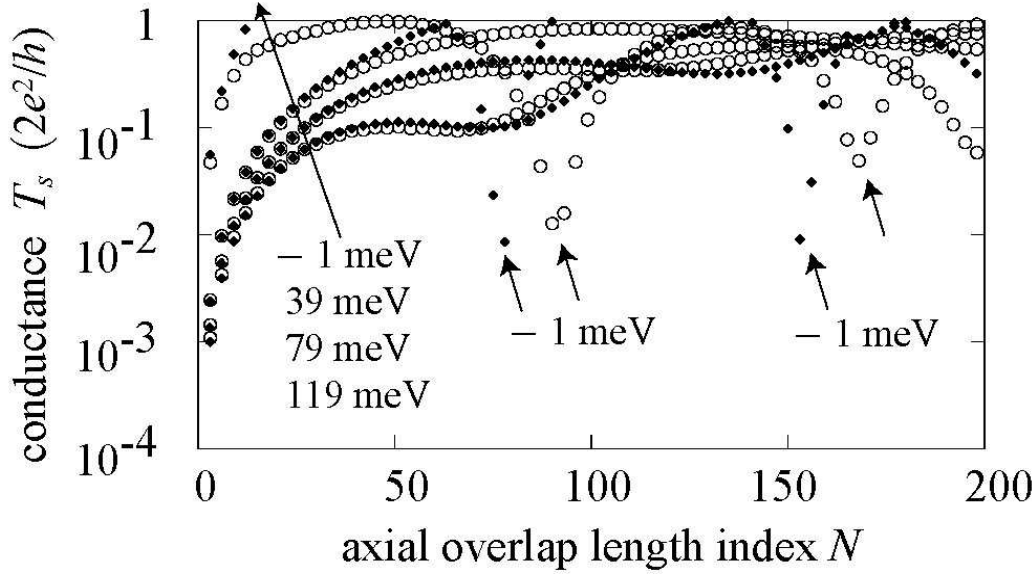}
\caption{The total transmission rates $T_s=T_1+T_2$, that is 
the Landauer's formula conductance in the $2\frac{e^2}{h}$ unit, 
as a function of the axial overlap length
index $N$ in case mod($N,3$)=0, $E=-1, 39, 79, 119$ meV
and $M=16$ (configuration B).
The circles and solid diamonds represent $T_s$
of the exact calculation and PF, respectively. 
}
\end{center}
\end{figure}

Figure 4 shows the sum of the transmission rates $T_s=T_1+T_2$,
that is, Landauer's formula conductance in the $2e^2/h$ unit as a function of the axial overlap length $N$ in case mod($N,3$)=0, $E=-1, 39, 79, 119$ meV and $M=16$ (configuration B).
The circles and solid diamonds represent $T_s$
for the exact calculation and PF, respectively.
The PF becomes ineffective with a large $N$ because the total interlayer interaction that is proportional to $N$, becomes too large to be regarded as a perturbation.
However, the PF works well in a finite range of $N$.
Figure 4 shows the effective $N$ range; the PF reproduces the exact $T_s$ in the range $N<9, 50, 100, 100$ when $E=-1, 39, 79, 119$ meV.
As $|E|$ increased, the effective $N$ range increased.
This $E$ dependence probably originates from the edge bands; when $|E|$ is sufficiently large, the gradient of the edge band is close to $\frac{\sqrt{3}}{2}ta$ demonstrating the applicability
of the sc-ANT theory.
When $E=-1 $ meV, the PF reproduces the dips in $T_s$ 
with an underestimation $N$.
$N=78, 153$ in the PF and $N=90, 168$ in the exact calculation.
When $T_s$ of the PF exceeds one, the exact $T_s$ 
reach near the maximum, i.e., one.
Similar effectiveness of PF was confirmed.
for mod$(N,3)$ =1,2 and configuration A (not shown in Figure).

\begin{figure}
\begin{center}
\includegraphics[width=0.9\linewidth]{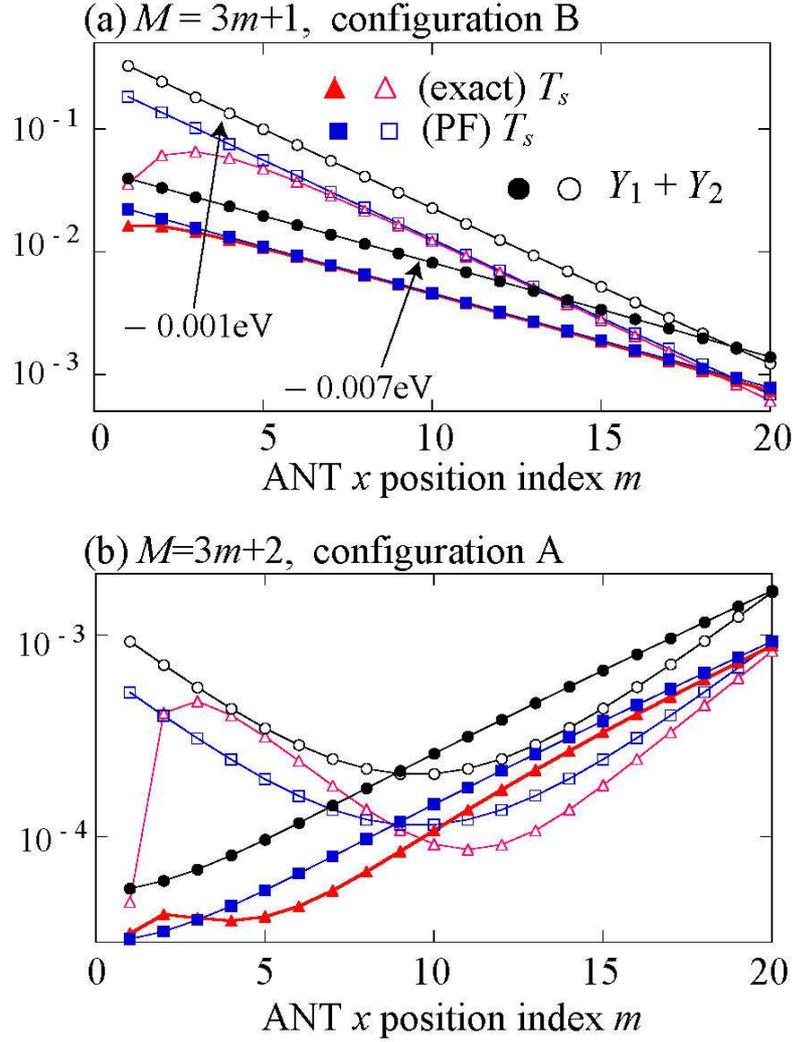}
\caption{(color online) Total transmission rate $T_s$ of the 
exact calculation (triangles) and PF calculations (squares)
as a function of the ANT $x$ position index $m$ in case $N=3$.
The circles represent $Y_1+Y_2$ calculated by Eq. (\ref{Y}).
The open and solid symbols
are the data from $E=-1$ meV and $E=-7$ meV,
respectively.
(a) Configuration B ($M= 3m+1$).
(b) Configuration A ($M= 3m+2$).
}
\end{center}
\end{figure}

Even when $|E|=1$ meV, the PF is effective with a small axial
overlap length $N=3$; thus, we choose $N=3$ for the following calculation.
A small overlap length $(N-2)a/2=a/2$ reduces the interlayer cohesion and allows for smooth sliding of the ANT on the ZGR.
Figure 5 (a)
shows the total transmission rate $T_s$ of the exact calculations (triangles) and the PF calculations (squares) as functions of the ANT $x$ position index $m$ in configuration B ($M=3m+1)$.
The circles represent $Y_1+Y_2$.
The open and solid symbols show the data for $E=-1$ meV and $E=-7$ meV,
respectively.
Figure 5 (b) is the same as Fig. 5 (a), except for the configuration 
changes from B to A.
The PF satisfactorily reproduced the variations in
 the exact $T_s$ with $m$; when $E \simeq 0$ and $(|\Delta k_1|-|\Delta k_2|)N \ll 1$, Eq. (\ref{PF}) proves that $T_s/(Y_1+Y_2) $  is independent of $m$, presenting a close relationship between $T_s$ and the ZGR wave function amplitude $|c'|^2$.
Figure 5 (a) shows the exponential localization of the B site in Fig. 3.
An increase in $|E|$ causes delocalization.
The $T_s$ and B-site amplitudes shown in Fig. 3 share the same slope on the semi-log scale.
In contrast, Fig. 5 (b) does not necessarily show B localization. In configuration A, the vertical bonds connect only the A sites of ZGR.
Although the B sites are also connected to ANT by oblique interlayer bonds, the factor $(-1)^{l-l'}$ in Eq. (\ref{Z}) strongly suppresses the contribution of B sites to Eq. (\ref{Z}) for configuration A.
In the open symbols ($E=-1$ meV) near the left edge $m < 10$, the A amplitude is negligible, and B-site localization emerges. 
However, as $m$ increases, the A-site amplitude increases and approaches the decaying B-site amplitude; thus, the open symbols increase with $m$.
In the solid symbols ($E= -7$ meV), the difference between the A and B sites decreases, as shown in Fig. 3, and the suppression of the B site amplitude in Eq. (5) becomes more significant than that in the $E=-1 $ meV case.
Thus, the solid symbols directly reflect the A-site amplitude, which increases exponentially with $m$.

Despite the edge localization, the phase relation of the ZGR cancels the B-site decaying wave and enables the detection of 
 the A-site wave function growth from the edge to the center.
This growing signal can be detected by a conventional STM tip; however, this has yet to be reported.
Although the edge roughness may destroy the amplitude growth, moderate edge roughness does not alter the sublattice amplitude difference in the theoretical calculations of Refs. \cite{zig-9-10-discovery,zig-3-exp-theory,zig-4-exp-theory} . 
The thermal vibration and shift from the AB stacking break the relation $H^{(l.l+1)} = H^{(l,l-1)}$ and weaken the $(-1)^{l-l'}$ cancelation in Eq. (\ref{Z}).
These issues should be addressed in future studies.
The range of the vertical axis in Fig. 5 is also notable.
The minimum in Fig. 5 (a) is close to the maximum in Fig. 5 (b).
The ammeter range must be changed by several orders of magnitude to detect the increase in amplitude according to the interlayer configuration.
Fortunately, the ANT-ZGR junction prefers the AB stacking configuration. 
Sliding the ANT tip along the armchair edge ($y=0$) causes the stable configurations A and B to alternate.
This helps to regulate the ammeter range according to the configurations A and B.
The estimated barrier height is 4 meV. \cite{barrier}
Although a more realistic Hamiltonian and atomic structure could be necessary for quantitative analysis, the main result -- detecting the wave function growth from the edge to the center -- is independent of the details of the model.
As the overlap between the opposite decay components governs the spin coupling between the opposite edges, this detection presents important information on ZGR magnetism \cite{zig-9-12-discovery-magnetism,zig-9-9-magnetism,zig-9-13-magnetism,zig-9-14-magnetism}, which promotes the application of the ANT tip to the graphene system.

\end{document}